\documentclass[12pt]{article}
\usepackage{epsfig} 
\def\beq{\begin{equation}}
\def\eeq{\end{equation}}

\def\al{\alpha}
\def\bt{\beta}

\def\de{\delta}
\def\De{\Delta}

\def\ka{\kappa}

\def\te{\theta}
\def\La{\Lambda}
\def\lam{\lambda}

\def\ep{\epsilon}

\def\sq{\sqrt}

\def\l{\left (}
\def\r{\right )}

\def\fr{\frac}
\def\la{\label}
\def\hs{\hspace}
\def\vs{\vspace}

\def\ran{\rangle}
\def\lan{\langle}
\def\ov{\overline}
\def\tl{\tilde}
\def\tm{\times}

\begin{document}

\begin{flushright}
BA-04-09\\
HD-THEP-04-31\\
August 11, 2004 \\
\end{flushright}

\begin{center}
{\Large\bf    

Sparticle Masses, $\mu $ Problem and Anomaly Mediated \\

Supersymmetry Breaking}
\end{center}

\vspace{0.5cm}
\begin{center}
{\large 
{}~Qaisar Shafi$^{a}$\footnote{E-mail address: 
shafi@bartol.udel.edu}~~and~ 
{}~Zurab Tavartkiladze$^{b}$\footnote{E-mail address: 
z.tavartkiladze@thphys.uni-heidelberg.de} 
}
\vspace{0.5cm}

$^a${\em Bartol Research Institute, University of Delaware,
Newark, DE 19716, USA \\

$^b$ Institute for Theoretical Physics, Heidelberg University,
Philosophenweg 16, \\
D-69120 Heidelberg, Germany}


\end{center}
\vspace{0.6cm}

\begin{abstract}

Within the MSSM framework and with purely anomaly mediated supersymmetry
breaking the slepton masses turn out to be tachyonic. We resolve this
problem by introducing an anomaly free $U(1)$ gauge symmetry which provides
positive $D$-term contributions to sparticle masses squared that are flavor
conserving at one loop. Two realistic examples based on $SU(5)$ are
presented. 
With $U(1)$ spontaneously broken at a scale $\sim 10^{16}$~GeV, 
the right handed neutrinos acquire 
masses $\stackrel{<}{_\sim}10^{14}$~GeV. This breaking scale of 
$U(1)$ also plays an important role in the proposed resolution of 
the MSSM $\mu $ problem.

\end{abstract}


\vs{0.5cm}

\section{Introduction}

Despite its many attractive features the anomaly mediated supersymmetry
breaking (AMSB) scenario  \cite{ran} has one very serious shortcoming, 
namely within the
MSSM framework the slepton masses turn out to be tachyonic. A number
of attempts to resolve this problem have appeared in the 
\cite{PR}-\cite{gravcosm}.
Recent interest in this scenario is largely spurred by the fact that the
gravitino mass in this approach can be considerably larger than a  TeV, perhaps
even as large as 100 TeV or so. 
This can be helpful as far as gravitino
cosmology is concerned \cite{gher}. A gravitino with mass 
considerably greater than a TeV
can decay before nucleosynthesis, thereby evading the severe constraints
on the reheat temperature $T_r$ that follow from nucleosynthesis 
considerations \cite{conTr}. 
Indeed, it has been argued that for $m_{3/2} > 60$~TeV, the upper bound on
$T_r$ from nucleosynthesis effectively disappears \cite{gher}. 
A new bound on $T_r$ can
arise from considerations of the LSP, especially if the latter happens to be
the neutral wino. In this case $T_r$ is estimated to be 
$\stackrel{<}{_\sim } 10^{11}$~GeV, which
is still significantly higher than the bound
$T_r\stackrel{<}{_\sim } 10^{5}-10^{9}$~GeV
for $m_{3/2}\sim $~TeV in gravity  
mediated supersymmetry breaking models. This makes inflationary model
building and successful lepto-baryogenesis considerably easier.

In this paper we propose to eliminate the tachyonic slepton masses by
introducing an additional source of supersymmetry breaking for the MSSM fields
via a superheavy messenger sector, taking care that the unification of the
MSSM gauge couplings is preserved. An anomaly free $U(1)$ gauge symmetry is
introduced under which both messengers and the MSSM fields transform non-
trivially, and which gives rise to flavor universal 1-loop contributions to
the sfermion squared masses. These new contributions can easily overcome the
negative two loop AMSB contributions. Since the magnitude of the $U(1)$ gauge
coupling is undetermined, the gravitino mass $m_{3/2}$ could be as high as
$20-60$~TeV. We present two examples based on $SU(5)$. 
The first one is inspired
by the decomposition $E_6\to SO(10)\tm U(1)$ and leads to a universal positive
contribution for all sfermion squared masses, independent of flavor. In the
second example the $\ov{\bf 5}$ matter fields receive twice the 
contributions as    
the ${\bf 10}$ fields. The $U(1)$ symmetry plays other essential roles. 
For instance,
its breaking ensures generation of right handed neutrino masses of the desired
magnitude. It also plays an important role in the resolution of the 
MSSM $\mu $ problem , and its spontaneous breaking may be linked to 
supersymmetric hybrid inflation.

\section{Extended AMSB Scenario}

In the minimal version of AMSB \cite{ran}, the source of SUSY breaking is the
non-zero $F_{\phi }$ component of a compensator superfield
$\phi =1+\te^2F_{\phi }$, with $\lan F_{\phi } \ran =m_{3/2}$.
This causes the generation of  soft SUSY breaking terms 
for the MSSM gauginos and sparticles through 
one and two
loop contributions respectively. Namely, the gaugino 
masses at one loop level are \cite{ran}
\beq
M_{\lam a }(\mu )=\fr{\al_a (\mu )}{4\pi }b_am_{3/2}~,
\la{gaugino}
\eeq
while the two loop contributions to sfermion squared masses are
\beq
m_i^2(\mu )=\fr{2m_{3/2}^2}{(4\pi )^2}\sum_{a=1}^3c^a_ib_a\al^2_a(\mu )~,
\la{msquared}
\eeq
where, for MSSM, $(b_1, b_2, b_3)=(-\fr{33}{3}, -1, 3)$, $c^a_i>0$, and 
the masses and couplings are evaluated at scale $\mu $.
The mass of the gravitino $m_{3/2}$ can be in the
$20-60$~TeV range because of
suppression by loop factors $\sim \fr{\al_a}{4\pi }$. This  insures 
that the sparticle masses can be in the TeV region as required by
the gauge hierarchy problem.
Although the scenario looks very attractive, its minimal version is ruled out
because of the negative  squared masses for the sleptons ($b_1, b_2<0$). 
Thus, new contributions, especially for the sleptons,
are needed in order to make the AMSB scenario realistic.

\subsection{Generation of Soft Masses at One-Loop }

In order to cure the problem of tachyonic sleptons we introduce  
$n$ pair of vector like messenger superfields:
$(\ov{\Psi }+\Psi )_i$,  $i=1,\cdots , n$. In addition, we introduce  
$U(1)$ gauge symmetry under which the MSSM states and messengers
transform non trivially. From the multimessenger ($n>1$) sector and $U(1)$ 
factor, the sfermion masses$^2$ can obtain flavor universal 
1-loop contributions \cite{dvali, DG}. We will show that this gives an elegant 
possibility to avoid the shortcomings of the minimal AMSB scenario and 
to build realistic models.

The $D$-term corresponding to the $U(1)$ gauge factor is given by
\beq
D=Q_{\Psi }(\tl{\Psi }^{\dagger }_i\tl{\Psi }_i-
\tl{\ov{\Psi }}_i\tl{\ov{\Psi }}^{\dagger }_i)+Q_{f}\tl{f}^{\dagger }\tl{f}~,
\la{Du1}
\eeq
where $f$ denotes the MSSM states and $Q_{\Psi ,f}$ stand for the 
$U(1)$ charges of the corresponding  superfields.
We assume that the messenger sector feels SUSY breaking through 
the non-zero $F_X$ component of a chiral superfield $X$, 
which then gets transferred to the visible sector. The relevant
superpotential couplings for the messengers are
\beq
W(\Psi )=M_{ii}\ov{\Psi }^i\Psi^i+\lam_{ij}X\ov{\Psi }^i\Psi^j~,
\la{mesW}
\eeq
where, without loss of generality, we have chosen a basis in which
the matrix $M$ is diagonal with  positive elements, and we also
assume that the lowest component of $X$ has no VEV. 
The fermionic components from $\Psi_i, \ov{\Psi }_i$ have masses
$M_{ii}$, while their scalar partners are split due to SUSY breaking. 
The scalar $2n\tm 2n$ mass matrix is given by
\beq
\begin{array}{cc}
 & {\begin{array}{cc}
\hs{-0.1cm} \tl{\Psi }&\hs{0.22cm} \tl{\ov{\Psi }}^{\dagger }
\end{array}}\\ \vspace{2mm}
\hat{M}^2_{\Psi }=
\begin{array}{c}
 \tl{\Psi}^{\dagger }\\ \tl{\ov{\Psi }} 

\end{array}\!\!\!\!\!\! &{\left(\begin{array}{cc}
M^2~ &~ F_X^*\lam^{\dagger}
\\
F_X\lam ~ &~ M^2

\end{array}\!\right )}~.
\end{array}  
\la{leading}
\eeq
It is clear that even if the lowest component of $X$ has a VEV, 
it can be absorbed 
in $M$, and  one can then choose the basis in which $M$ is diagonal. 
It is essential, though, that the matrix $\lam $ which couples with $F_X$
in (\ref{mesW}) is non-diagonal. Performing the
transformation $U\hat{M}^2_{\Psi }U^{\dagger }=\hat{M}^{'2}_{\Psi }$
with
\beq
\begin{array}{cc}
 & {\begin{array}{cc}

\end{array}}\\ \vspace{2mm}
U=
\begin{array}{c}

\end{array}\!\!\!\!\!\! &{\fr{1}{\sq{2}}\left(\begin{array}{cc}
{\bf 1}_{n\tm n} &~ -{\bf 1}_{n\tm n}
\\
{\bf 1}_{n\tm n}~ &~ {\bf 1}_{n\tm n}

\end{array}\!\right )}~~~~~{\rm and }~~~~
(\tl{\Psi }^{\dagger }~,~\tl{\ov{\Psi }}) =
(\Phi^{\dagger }~,~\ov{\Phi }) U~,
\end{array}  
\la{UPhi}
\eeq
we find
\beq
\begin{array}{cc}
 & {\begin{array}{cc}
\hs{-0.1cm} {\Phi }&\hs{2.6cm} {\ov{\Phi }}^{\dagger }
\end{array}}\\ \vspace{2mm}
\hat{M}^{'2}_{\Psi }=
\begin{array}{c}
 {\Phi }^{\dagger }\\ {\ov{\Phi }} 

\end{array}\!\!\!\!\!\! &{\left(\begin{array}{cc}
M^2-({\cal F}+{\cal F}^{\dagger })~ &~ {\cal F}^{\dagger }-{\cal F}
\\
{\cal F}-{\cal F}^{\dagger }~ &~ M^2+{\cal F}+{\cal F}^{\dagger }

\end{array}\!\right )}~,
\end{array}  
\la{M1}
\eeq
where ${\cal F}=\fr{1}{2}F_X\lam $.
In the basis (\ref{UPhi})  the $D$-term of eq. (\ref{Du1}) becomes:
\beq
D=Q_{\Psi }(\ov{\Phi }_i\Phi_i +
{\Phi }^{\dagger }_i\ov{\Phi }^{\dagger }_i)
+Q_{f}\tl{f}^{\dagger }\tl{f}~.
\la{1Du1}
\eeq
From the $D^2$ term in the Lagrangian there exists an interaction term 
$Q_f\tl{f}^{\dagger }\tl{f}(\ov{\Phi }_i\Phi_1+{\rm h.c.})$, and 
with a non-vanishing insertion between $\Phi $ and $\ov{\Phi }$ 
[${\cal F}-{\cal F}^{\dagger }$ entry in matrix (\ref{M1})], the 
sfermions $\tl{f}$  acquire one loop masses \cite{dvali, DG}:
\beq
\tl{m}^2_{{f}}(1\hs{-0.15cm}-\hs{-0.1cm}{\rm loop})=
Q_f\fr{\tl{\al }}{4\pi }\La^2 ~,
\la{1lsoft2}
\eeq
where $\tl{\al }$ is the $U(1)$ `fine structure' coupling and 
the scale $\La $  depends on $M_{ii}$
and the couplings $\lam_{ij}$ (${\cal F}_{ij}$) \cite{DG}
\beq
\La^2=2{\rm Tr}\left [Q_{\Psi }\hs{-0.2cm}\sum_{i,j=1}^n\hs{-0.2cm}
\fr{|{\cal F}_{ji}|^2-|{\cal F}_{ij}|^2}{(M_{ii})^2}
f\l \fr{M_{jj}^2}{M_{ii}^2}\r \right ] ~,
~~~
{\rm with}~~~f(x)=\fr{2}{1-x}+\fr{1+x}{(1-x^2)}\ln x~.
\la{LamScale}
\eeq 
Note that for this one loop contribution to be non-zero, it is
essential that $Q_f\neq 0$ and the matrix 
${\cal F}$($=\fr{1}{2}F_X\lam $) is neither hermitian nor symmetric.

The result in (\ref{1lsoft2}) can be interpreted as the generation of a non 
zero VEV of the $D$-term, $\lan D\ran =\fr{\tl{\al }}{4\pi }\La^2$. Indeed, 
the messengers which couple to non-zero $F_X$, dynamically generate 
$\lan D\ran$. This can be seen by performing the shift 
$D\to D+\lan D\ran $ in (\ref{1Du1}) and  squaring, and we see that the 
soft term 
$\tl{m}^2_{{f}}=Q_f\lan D\ran $ emerges. In other words, since there 
are couplings $\int d^4\te f^{\dagger }e^{Q_fV}f$ in the 
K\"ahler potential, the 
contribution $V=\te^2\bar \te^2 \lan D\ran +\cdots $ induces soft mass 
squared terms. Since each generation of particles with identical
SM quantum numbers carry the same
$U(1)$ charge, the universality of sparticle 
masses is  ensured at 1-loop. This remains intact under renormalization 
since $\tl{m}_f\propto \lan D\ran $ belongs to the solution of RGE 
trajectory \cite{carena, hamed}. Thus, the FCNC are naturally suppressed.
Note that a potential mixing of the messengers with the MSSM
matter fields which could lead to large flavor violation is 
readily avoided because of the MSSM `matter' parity.

With $\lam_{ij}\sim 1$ and $M_{ii}$ all of comparable magnitude, 
we have $\La \sim \fr{F_X}{M_{ii}}$. With $F_X=m_{3/2}\ov{M}$
and $\tl{\al }\sim 1/25$ (this is a reasonable value especially if $U(1)$
arises from some GUT unifying the MSSM interactions at 
scales$\sim 10^{16}$~GeV)
we should have $M_{ii}\sim (3-6)\cdot \ov{M}$ in order to get
$\tl{m}_f\sim 1$~TeV (with the gravitino mass $m_{3/2}\simeq 20-60$~TeV). 
To estimate $\ov{M}$ note that the operators
$\int d^4\te \fr{X^{\dagger }X}{M_{Pl}^2}f^{\dagger }f$  
cause flavor violating contributions that are harmless if 
$\ov{M}\stackrel{<}{_\sim }3\cdot 10^{14}$~GeV. 
Thus, an order of magnitude estimate of the messenger masses is 
$M_{ii}\stackrel{<}{_\sim }10^{15}$~GeV. 
This bound can be modified if $\lam_{ij}$ are not of order unity, in 
which case the  $M_{ii}$'s can have different values.
A nice feature of this scenario is that there is 
just one additional parameter $\tl{\al }\La^2$ which contributes 
to the sparticle masses (at one loop).

The messengers also contribute to gaugino masses so that the 1-loop
expression for the latter is given by 
\beq
M_{\lam a }(\mu )=\fr{\al_a (\mu )}{4\pi }(b_a-nA\De b_a)m_{3/2}~,
\la{finMg}
\eeq
where $\De b_a$ comes from the messengers
(for a pair $\ov{\bf 5}+{\bf 5}$ of $SU(5)$, $\De b_a=-1$) and, 
for simplicity,
it is assumed that all messengers have comparable masses and
$F_X/M_{ii}=Am_{3/2}$.
The 2-loop contribution to the sparticle masses also get modified 
and is given by 
$$
m_i^2(2\hs{-0.15cm}-\hs{-0.1cm}{\rm loop})=
\fr{2m_{3/2}^2}{(4\pi )^2}\sum_{a=1}^3c^a_ib_a\left [\al^2_a(\mu )-
A\al^2_a(\mu )\fr{n\De b_a}{b_a}
+A^2\l \al^2_a(\mu )-\al^2_a(M_{ii})\r 
\fr{(n\De b_a)^2}{b_a^2}\right.
$$
\beq
\left. -A\l \al^2_a(\mu )+A\al^2_a(M_{ii}) \r 
\fr{n\De b_a}{b_a}
\right ]~,
\la{2lFinMs}
\eeq 
(with this expression one can recover the result of \cite{PR} 
for concrete case with $A=-1$ and $n\De b_a\to -N$).

It is natural to expect that the 1-loop contribution (\ref{1lsoft2})
dominates over the 2-loop AMSB contributions, so that the problem 
with tachyonic slepton masses can be easily avoided\footnote{One might ask 
whether the additional two loop contributions are sufficient
to overcome the negative 2-loop AMSB contributions , in which case the
1-loop contributions to the squared sparticle masses could be eliminated
by invoking a suitable 'messenger parity' \cite{dvali}. 
In the next section we present two examples based on $SU(5)$ with three 
families of ${\bf 5}+ \ov{\bf 5}$ messengers and,
as shown in \cite{PR}, this is not sufficient to eliminate the 
tachyonic slepton masses}.
Therefore, we propose that
the contributions (\ref{1lsoft2}) provide positive squared masses for
all the sfermions.
In particular, the U(1) charges of $l$ and $e^c$ 
should have the same sign (this is not case with $U(1)_{B-L}$.
(For attempts at resolving the problem with this abelian factor together with
other contributions see refs. \cite{jack, hamed, kit}). 
We present below two new examples within the $SU(5)$ framework 
which provide a successful realization of the mechanism presented in 
this section.

\section{Models with an Additional $U(1)$ Symmetry}

\subsection{Model A}

Our first model is inspired by the grand unified group $E_6$ \cite{E6}
with $U(1)$ corresponding to the decomposition
$E_6\to SO(10)\tm U(1)$. In an $SU(5)$ setting the anomaly free content 
per generation is
$$
{\bf 10}_1+\ov{\bf 5}_1~~~~~~~~~~~~({\rm Chiral~ states~ of~ MSSM})
$$
$$
\ov{\bf 5}_{-2}+{\bf 5}_{-2}~~~~~~~~~~~~~~~~~~~({\rm Messengers})
$$
\beq
{\bf \nu }^c_{1}+{\bf N}_4~~~~~~~~~~~~~~~~~~({\rm MSSM~ singlets})~,
\la{Qmodel1}
\eeq
where the subscripts label the $U(1)$ charge.
The three $(\ov{\bf 5}+{\bf 5})_{-2}$ families  which are crucial 
for anomaly 
cancellation can be used as messengers. That is,
the number of messengers $n=3$ coincides with the number of 
MSSM quark-lepton generations. 
The messenger superpotential couplings are
\beq
W_1=(\de_{ij}m_{ii}+
\hat{\lam }_{ij}\fr{X}{\phi M_*})\chi_4\ov{\bf 5}_{-2}^i{\bf 5}_{-2}^j~,
\la{mesW1}
\eeq
($X$ and $\phi $ have zero $U(1)$ charge)
where $m_{ii}$, $\hat{\lam }_{ij}$ are dimensionless couplings, $M_*$
denotes the cut-off, and
$\chi_4$ is an $SU(5)$ singlet superfield.
The VEV of its scalar 
component breaks $U(1)$ and also generates masses for the messengers.
With
$M_{ii}=\lan \chi_4\ran m_{ii} $, 
$\lam_{ij}=\hat{\lam }_{ij}\lan \chi_4\ran /M_*$,
the superpotential (\ref{mesW1}) has the ingredients needed
for sfermion mass$^2$ generation at 1-loop. 
With $Q_{\bf 10}=Q_{\ov{\bf 5}}$, according to (\ref{1lsoft2})
we have the prediction:
\beq
\tl{m}^2_{\ov{\bf 5}}(1\hs{-0.15cm}-\hs{-0.1cm}{\rm loop})=
\tl{m}^2_{\bf 10}(1\hs{-0.15cm}-\hs{-0.1cm}{\rm loop})\equiv \ov{m}^2~.
\la{1lsoftM1}
\eeq
Thus, we have a positive universal (1-loop) contribution for
all the sfermions' squared masses.
Radiative electroweak (EW) breaking and sparticle spectroscopy
with this type of asymptotic relation has been discussed in the
literature \cite{gher}. We see that it arises within a rather 
simple example of SUSY $SU(5)$ supplemented by a suitable 
anomaly free $U(1)$ gauge symmetry.

Note that since 
for each $SU(5)$ multiplet $\Psi $ we have ${\rm Tr}Y_{\Psi }=0$,
a $D$-term for $U(1)_Y$ is not generated at one loop (see discussion 
in \cite{DG}). Therefore, only the $D$-term of $U(1)$ plays a 
role in generating 1-loop sparticle masses.

For a more precise estimate of masses we should take into account the
2-loop contributions in (\ref{2lFinMs}). As an example we 
take $m_{3/2}=50$~TeV, 
messenger masses $M_{ii}\approx 10^{15}$~GeV 
[i.e. $A\approx -1/6$ in (\ref{2lFinMs})] 
and $\ov{m}=500$~GeV in 
(\ref{1lsoftM1}).
We then obtain the following sparticle mass spectra at $1$~TeV scale:
\beq 
0.99 m_{\tl{u^c}}\simeq 0.98m_{\tl{d^c}}\simeq 2.31m_{\tl{l}}
\simeq 2.22m_{\tl{e}^c}\simeq m_{\tl{q}}~,~~~~~
m_{\tl{q}}\simeq 1.02~{\rm TeV}~.
\la{spectra1}
\eeq
{}For this analysis we have ignored the Yukawa couplings which 
may be relevant for the third generation
(small or moderate values of the MSSM parameter 
$\tan \bt $ are preferred by considering the
Yukawa sector; see below).
Note that the left and right handed sleptons of the first two generations
are quasi-degenerate in mass  to within $4\%$.
The gaugino masses are determined according to (\ref{finMg}),
\beq
M_{\lam_1}\simeq 2.48 M_{\lam_2}\simeq 0.5M_{\lam_3}~,~~~~~
M_{\lam_1}\simeq 488~{\rm GeV}~.
\la{Mgspectra1}
\eeq

In the higgs sector, together with $H({\bf 5}_{-2})$,  
$\ov{H}(\ov{\bf 5}_2)$ which contain the MSSM doublets $h_u$, 
$h_d$ respectively, we have the singlets $\ov{\chi }_{-4}$, $\chi_{4}$
whose VEVs break the $U(1)$
symmetry. The Yukawa superpotential couplings responsible 
for the generation of quark and lepton masses are
\beq
W_{1Y}={\bf 10}_1\cdot {\bf 10}_1\cdot H({\bf 5}_{-2})+
\fr{\ov{\chi }_{-4}}{{M}'}
{\bf 10}_1\cdot \ov{\bf 5}_1\cdot \ov{H}({\ov{\bf 5}}_{2})~,
\la{WY1}
\eeq
where the generation indices are suppressed and the cut off $M'$ here 
(and below) is expected to be comparable to $M_{Pl}$. 
If $\lan \ov{\chi }_{-4}\ran /{M'}\ll 1$,the $b, \tau $
Yukawa couplings are sufficiently small and we expect that $\tan \bt $
is not too large.

The 1-loop soft mass$^2$ for  the MSSM higgs doublet $h_u$ is negative 
(because of its negative $U(1)$ charge).
Including the $\mu $ and 
$B\mu $-terms the $h_u-h_d$ mass matrix 
(before loop corrections from the top and stop are included) is given by
\beq
\begin{array}{cc}
 & {\begin{array}{cc}
\hs{-0.1cm} h_u\hs{1.2cm} &\hs{1.2cm}h_d^{\dagger }
\end{array}}\\ \vspace{2mm}

\begin{array}{c}
 h_u^{\dagger }\\ h_d

\end{array}\!\!\!\!\!\! &{\left(\begin{array}{cc}
\mu^2-2\ov{m}^2-m_0^2~ &~ B\mu
\\
B^*\mu~ &~ \mu^2+2\ov{m}^2-m_0^2

\end{array}\!\right )}~,
\end{array}  
\la{hudMass}
\eeq  
where $m_0^2\simeq (240)^2~{\rm GeV}^2$ denotes the (negative) 
2-loop contribution.
The generation of $\mu $ and $B\mu $-terms will be discussed in section 4.

Some comments about the $U(1)$ symmetry breaking are in order. 
The VEVs of the lowest components of $\ov{\chi }_{-4}, \chi_4$ states are 
necessary for the generation of messenger masses. However, the scale of $U(1)$ 
breaking is not yet constrained, because the messenger masses 
also involve the Yukawa couplings  $m_{ii}$. On the other hand, the 
MSSM singlet states $\nu^c_1$ in 
(\ref{Qmodel1}) can be used as  right handed neutrinos whose Majorana 
masses are related to the $U(1)$ breaking scale. The Dirac Yukawa 
couplings are
$\ov{\bf 5}_1H_{-2}\nu^c_1$. For the Majorana masses of $\nu^c_1$ we 
introduce the scalar singlets $\ov{\xi }_{-1}+\xi_1$, and the masses arise
through the coupling
$(\nu^c_1\ov{\xi }_{-1})^2/M_{Pl}$.  To 
accommodate the atmospheric neutrino data 
one needs $\lan \ov{\xi }\ran \sim 10^{16}$~GeV, which gives
$m_{\rm atm }\sim \lan h_u^0\ran^2M_{Pl}/\lan \ov{\xi }\ran^2\sim 
0.1$~eV. The states $N_4$ do not play any role here 
in generating the light 
neutrino masses. They decouple by acquiring superheavy masses
via the couplings 
$(N_4\ov{\chi }_{-4})^2/M_{Pl}$.  
The presence of right handed neutrinos will induce lepton flavor violating
effects that are suppressed by an additional loop factor compared to the
dominant flavor universal contributions proportional to 
$\lan D\ran $. The soft SUSY
breaking parameters, to leading order, still belong to the UV insensitive
anomaly mediated trajectory \cite{gravcosm}.

We see that the $U(1)$ symmetry breaking plays an essential role 
for the generation of 
messenger masses as well as for realizing masses appropriate for the 
description of neutrino oscillations. 
In section 4, the importance of $U(1)$ in resolving the MSSM $\mu $ problem
will be discussed.

\subsection{Model B}

The second $SU(5)$ scenario has the following `matter' content:
$$
{\bf 10}_1+\ov{\bf 5}_2~~~~~~~~~~~~({\rm Chiral~ states~ of~ MSSM})
$$
$$
{\bf 5}_{-2}+\ov{\bf 5}_{-3}~~~~~~~~~~~~~~~~~~~({\rm Messengers})
$$
\beq
{\bf N}_5~~~~~~~~~~~~~~~~~~({\rm MSSM~ singlet})~,
\la{Qmodel2}
\eeq
with the messenger superpotential given by
\beq
W_2=(\de_{ij}m_{ii}+
\hat{\lam }_{ij}\fr{X}{\phi M_*})\chi_5\ov{\bf 5}_{-3}^i{\bf 5}_{-2}^j~,
\la{mesW2}
\eeq
where the singlet superfields $\chi_5, \ov{\chi }_{-5}$ 
are needed to break $U(1)$.

In 1-loop approximation the sparticle masses are predicted to be
\beq
\tl{m}^2_{\ov{\bf 5}}(1\hs{-0.15cm}-\hs{-0.1cm}{\rm loop})=
2 \cdot \tl{m}^2_{\bf 10}(1\hs{-0.15cm}-\hs{-0.1cm}{\rm loop})
\equiv \ov{m}^2~.
\la{1lsoftM2}
\eeq
The Yukawa superpotential involves the following couplings
\beq
W_{2Y}={\bf 10}_1\cdot {\bf 10}_1\cdot H({\bf 5}_{-2})+
\fr{\ov{\chi }_{-5}}{M'}{\bf 10}_1\cdot \ov{\bf 5}_2\cdot \ov{H}({\ov{\bf 5}}_{2})~,
\la{WY2}
\eeq
so that $\tan \bt $ is also not large for this
model. 

Taking into account the 2-loop contributions in  (\ref{2lFinMs}), 
and with $\ov{m}=500$~GeV, $m_{3/2}=50$~TeV, 
$M_{ii}\approx 10^{15}$~GeV 
[i.e. $A\approx -1/6$ in (\ref{2lFinMs})], 
the sparticle mass spectra is given by
\beq
0.99 m_{\tl{u^c}}\simeq 0.88m_{\tl{d^c}}\simeq 1.53m_{\tl{l}}
\simeq 2.22m_{\tl{e}^c}\simeq m_{\tl{q}}~,~~~~~
m_{\tl{q}}\simeq 1.02~{\rm TeV}~.
\la{spectraB}
\eeq
This is different from the one obtained for Model {\bf A} 
due to the asymptotic relation (\ref{1lsoftM2})
(there is no mass degeneracy any more between left and right handed
sleptons). 
The gaugino masses are the same  as for 
Model {\bf A} (eq. (\ref{Mgspectra1})). 
The $h_u-h_d$ mass matrix for Model {\bf B} coincides with (\ref{hudMass}).

\section{$\mu $ Problem and $U(1)$ Symmetry}

The presence of the $U(1)$ symmetry can be exploited to yield a solution 
of the $\mu $ problem in models with AMSB, along the lines proposed in
\cite{DLS} for gravity mediated SUSY breaking. To achieve the breaking of 
$U(1)$ with a renormalizable superpotential, a gauge singlet chiral 
superfield
$S$ is introduced. The singlet $S$ triggers the $U(1)$ breaking, and 
its scalar and auxiliary components acquire VEVs after SUSY breaking 
which generate the $\mu $ and $B\mu $-terms of the MSSM. Last but not 
least, $S$ can play the role of the inflaton field in supersymmetric 
hybrid inflation \cite{DSS}.

The superpotential 
\beq
W_S=\ka S(\ov{\chi }\chi -\phi^2M^2)~,
\la{WS}
\eeq
involves the pair $\ov{\chi}, \chi $ (needed for $U(1)$ breaking) and, 
in the unbroken SUSY limit (i.e. $\lan \phi \ran =1$, $F_{\phi }=0$), 
$\lan S\ran =0$ (the dimensionless coupling $\ka >0$ without loss 
of generality). The presence of $\phi^2$
in (\ref{WS}) is required for the correct Weyl weight of $W_S$, with all 
superfields having the canonical Weyl weights($=1$). 
Note that we have also employed a $U(1)$ $R$-symmetry such that $W_S$ and
$S$ have a unit $R$-charge.
With SUSY unbroken, the solution
$\lan S\ran =0, |\lan \ov{\chi }\ran |=|\lan \chi \ran |=\phi M$ satisfies
$F$ and $D$ flatness conditions. Substituting $\phi =1+\te^2m_{3/2}$ 
the scalar potential derived from (\ref{WS}) 
is given by
\beq
V=\ka^2|\ov{\chi }\chi -M^2|^2+\ka^2(|\chi |^2+|\ov{\chi }|^2)
-(2\ka M^2m_{3/2}S+{\rm h.c.})
\la{VS}
\eeq
The soft SUSY breaking term in (\ref{VS}) shifts the VEVs:
\beq
\lan S\ran \simeq \fr{m_{3/2}}{\ka }~,~~~
\lan \ov{\chi }\ran = \lan \chi \ran \simeq M(1-\fr{m_{3/2}^2}{2\ka^2M^2})~,
\la{VEVshift}
\eeq
and also the corresponding $F$-terms
\beq
F_S\simeq -\fr{m_{3/2}^2}{\ka }~,~~~~~
F_{\chi }=F_{\ov{\chi }}\simeq m_{3/2}M~.
\la{Fshift}
\eeq
Note that although SUSY is broken we still have 
$|\lan \ov{\chi }\ran |=|\lan \chi \ran |$ and therefore there are
no additional contributions to $\lan D\ran $. Also, 
$\ov{\chi }, \chi $ sit on the AMSB trajectory
(i.e. $\fr{F_{\ov{\chi }}}{\lan \ov{\chi }\ran }=
\fr{F_{\chi }}{\lan \chi \ran }=m_{3/2}$) and therefore do not 
contribute, even not via loops, to the soft masses.
With $\lan S\ran =m_{3/2}/\ka $, the superpotential coupling
$\ka_hSh_uh_d$ gives $\mu \sim (\ka_h/\ka)m_{3/2}$, where $\ka_h$
is a dimensionless coupling. {}For $m_{3/2}\sim 50-100$~TeV, we need
$\ka_h\sim 10^{-2}\ka $ to generate a $\mu $-term$\sim $~TeV. 

Let us now 
recall that in sect. 3.1 we employed the $SU(5)$ singlets
$\ov{\xi }, \xi $ to generate masses for the right handed neutrinos. 
If the gauge invariant combinations 
$h_uh_d$ and $\ov{\xi }\xi $ transform under some discrete symmetry 
such  that $\ov{\xi }\xi \cdot h_uh_d$ is invariant, then the operator
\beq
W_{\mu }=\ka_h\fr{\ov{\xi }\xi }{(\phi M_{Pl})^2}Sh_uh_d~,
\la{Wmu}
\eeq
with
$\lan \ov{\xi }\ran =\lan \xi \ran \equiv \ep M_{Pl}$ gives
\beq
\mu\simeq \ep^2\fr{\ka_h}{\ka }m_{3/2}~.~~~~
\la{muBmu}
\eeq
With $\ka_h/\ka \sim 1$, $\ep \sim 0.1$ is sufficient to guarantee
$\mu \sim  $~TeV. 

The $B\mu $ term turns out to be somewhat large 
($\sim \mu m_{3/2}$) and some fine tuning may be necessary to implement
radiative electroweak breaking. 
Namely, from (\ref{Wmu}),
taking into account (\ref{VEVshift}), (\ref{Fshift}), we have
\beq
B\mu =-\mu m_{3/2}\l 3-2\fr{F_{\xi }}{\lan \xi \ran m_{3/2}}\r ~,
\la{Bmu}
\eeq 
where $F_{\xi }$ denotes the VEV of the auxiliary component of 
$\xi $ ($\ov{\xi }$) and its magnitude can be expected to be of
order $F_{\xi }\sim \lan \xi \ran m_{3/2}$.
With
$F_{\xi }/(\lan \xi \ran m_{3/2})\simeq 3/2+{\cal O}(\fr{\mu }{m_{3/2}})$ 
we can arrange that $B\mu \sim \mu^2$.

The superpotential (\ref{WS}) was extensively used for building inflationary 
scenarios within various realistic models \cite{DSS}-\cite{pref}. 
Indeed, for successful 
inflation including leptogenesis, $\ka $ near $10^{-2}$ is 
preferred \cite{pref}, in which case
$\ep \sim 10^{-2}$. This corresponds to 
$\lan \ov{\xi }\ran =\lan \xi \ran \sim 10^{16}$~GeV,  which is also
preferred  for realizing the observed neutrino 
masses (see sect. 3.1).

\section{Conclusions}

An anomaly free $U(1)$ gauge symmetry is employed to solve the problem of
tachyonic slepton masses encountered in models with anomaly mediated
supersymmetry breaking. Two simple examples based on $SU(5)$ are presented.
With $U(1)$ spontaneously broken at a scale 
$M\sim 10^{16}$~GeV, the right handed
neutrinos acquire masses$\stackrel{<}{_\sim }10^{14}$~GeV, which is 
suitable for realizing
the correct light neutrino masses needed for neutrino oscillations
and for implementing leptogenesis. A
mechanism for resolving the MSSM $\mu $ problem is discussed in which the scale
$M$($\sim 10^{16}$~GeV) also plays an essential role. 
Finally, the breaking of $U(1)$ can be linked
to hybrid inflation such that $\de T/T$ is proportional to $(M/M_{Pl})^2$.
Thus, $U(1)$ in our approach
plays an essential role in the construction of realistic models
utilizing anomaly mediated supersymmetry breaking.     

\vs{0.5cm}

\hs{-0.7cm}{\bf Acknowledgments}

\vs{0.2cm} 
\hs{-0.7cm}Q.S. would like to acknowledge the hospitality provided 
by the Institut f\"ur Theoretische Physik in Heidelberg, especially 
Michael Schmidt and Christof Wetterich, and he also thanks the Alexander 
von Humboldt Stiftung.  This work is supported by 
NATO Grant PST.CLG.977666 and
by DOE under contract DE-FG02-91ER40626.


\bibliographystyle{unsrt}

\end{document}